\documentclass[epsfig,12pt]{article}
\usepackage{graphicx}
\usepackage{amsfonts}
{\catcode `\@=11 \global\let\AddToReset=\@addtoreset}

\def\greaterthansquiggle{\raise.3ex\hbox{$>$\kern-.75em\lower1ex\hbox{$\sim$}}}
\def\lessthansquiggle{\raise.3ex\hbox{$<$\kern-.75em\lower1ex\hbox{$\sim$}}}

\newcommand{\beq}{\begin{equation}}
\newcommand{\eeq}{\end{equation}}
\newcommand{\beqa}{\begin{eqnarray}}
\newcommand{\eeqa}{\end{eqnarray}}
\newcommand{\beqan}{\begin{eqnarray*}} 
\newcommand{\eeqan}{\end{eqnarray*}}
\newcommand{\ba}{\begin{array}}
\newcommand{\ea}{\end{array}}

\newcommand{\A}{{\mathcal A}}
\newcommand{\B}{{\mathcal B}}
\newcommand{\Oo}{{\mathcal O}}

\newcommand{\M}{{\mathcal M}}

\newcommand{\GG}{{\mathcal G}}

\def\tr{\mathrm{tr\,}}

\newcommand{\dfrac}{\displaystyle \frac}
\newcommand{\dint}{\displaystyle \int}
\newcommand{\dsum}{\displaystyle \sum}

\normalsize
\begin{document}

\begin{titlepage}
\thispagestyle{empty}

\title{Quantum gauge fields and flat connections in
2-dimensional BF theory
\\[36pt]}

\author{Anton Alekseev \\
{\small\it Section de math\'ematiques, Universit\'e de Gen\`eve}\\ [-3pt]
{\small\it 2-4 rue du Li\`evre, c.p. 64, 1211 Gen\`eve 4, Switzerland} \\[16pt]
and \\[16pt]
Nevena Ilieva\\
{\small\it Institute for Nuclear Research and Nuclear Energy}\\[-3pt]
{\small\it 72, Blvd. Tsarigradsko Chaussee, Sofia 1784, Bulgaria}
}

\date{{\small\sl\today}}
\maketitle

\begin{abstract}

The 2-dimensional BF theory is both a  gauge theory and a topological 
Poisson $\sigma$-model corresponding to a linear Poisson bracket. In \cite{To1},
Torossian discovered a connection which governs correlation functions
of the BF theory with sources for the $B$-field. This connection is flat,
and it is a close relative of the KZ connection in the WZW model.
In this paper, we show that flatness of the Torossian connection
follows from (properly regularized) quantum equations of motion
of the BF theory.

\vspace{60pt}

\begin{quotation}
\end{quotation}

\end{abstract}

\vfill
\end{titlepage}

\setcounter{page}{2}

\section{Introduction}

The 2-dimensional BF theory is a an interesting example of a model which is
at the same time a gauge theory and a (topological) Poisson $\sigma$-model
corresponding to a linear Poisson bracket. Hence, we have an interesting opportunity
to compare two different approaches to quantization of the model.

As a Poisson $\sigma$-model, the BF theory gives rise to a star product 
on the dual space of a Lie algebra $\GG$ (see \cite{CatFel}). The 
Kontsevich approach to quantization is to fix the gauge and to study the
Feynman graphs of the model \cite{Kont}. In this context, Torossian
\cite{To1} discovered a very interesting flat connection which governs
the behavior of correlation functions of exponentials of the $B$-field.
This connection is a close relative of the Knizhnik-Zamolodchikov connection
\cite{KZ} in the WZW model.

Our aim in this paper is to better understand the origin of the Torossian
connection from the point of view of gauge theory. To this end, we consider the
BF theory with source terms for the $B$-field placed at the points
$z_1, \dots, z_n$, and we study the expectations of the quantum gauge field $\A$
and of the quantum $\B$-field. In terms of Feynman diagrams, we obtain tree contributions
for the  field $\A$ and one-loop (wheel) contributions for $\B$. Quantum fields $\A$ and $\B$ satisfy
quantum equations of motion which actually coincide with the classical ones.

In order to control the behavior of correlators, we need to specify the quantum gauge field $\A$ at the points $z_1, \dots, z_n$ where the source terms are located. Since $\A$ diverges at these points, we regularize it by subtracting the pole. At the level of Feynman diagrams, this corresponds to excluding one particular length-one tree from summation 
(the choice of this short tree depends on the point $z_i$). The set of regularized
values $\A^{reg}(z_1), \dots, \A^{reg}(z_n)$ form a connection $\mathbb{A}$ on the space
of configurations of points $z_1, \dots, z_n$. This connection governs the behavior of correlation functions, and it  takes values in the Lie algebra of 
vector fields on $n$ copies of $\GG$.

It turns out that the connection ${\mathbb A}$ is flat \cite{AT0906}. We explain the flatness of ${\mathbb A}$ as a consequence of the quantum equations of motion for the 
fields $\A$ and $\B$.

The paper is organized as follows. In Section 2, we briefly recall the basics of the BF theory,
the Feynman diagrams and classical and quantum equations of motion. In Section 3,
we study the dependence of the correlation functions on the sources, introduce the 
regularized gauge field and consider the flatness property of the connection ${\mathbb A}$. 

\medskip
\noindent{\bf Acknowledgements:}
We thank T. Strobl and C. Torossian for useful discussions and remarks. We are grateful to the 
International Erwin Schr\"odinger Institute for Mathematical Physics 
for the stimulating atmosphere.
This research was supported in part by the grants
200020-121675 and 200020-120042 of the Swiss National Science Foundation.

\bigskip

\section{Classical and quantum BF theory}

\subsection{Classical action and equations of motion}

Topological field theories \cite{WittTFT} (see \cite{BlauPhysRep} for a review) were introduced about 20 years ago as a novel class of field theories whose partition functions are independent of the metric. 
In particular, the BF theory is a  topological gauge theory which can be defined in any dimension. 
Let $G$ be a connected Lie group, $\GG$ its Lie algebra, and denote by $\tr (ab)$ 
an invariant scalar product on $\GG$ (for instance, the Killing form if $G$ is semisimple).
For $\M$ an oriented manifold of dimension $n$ (the space-time of the model) 
and $P$ a principal $G$-bundle over $\M$,
fields of the BF theory are the gauge field $A$ on the bundle $P$ and the
$\GG$-valued $(n-2)$-form $B$.  The action is given by
\beq\label{BF0}
S_{BF} = \tr \int BF, \qquad F = dA + \frac{1}{2} [A, A].
\eeq
Its quadratic part is of the first order in derivatives, so the theory has no physical degrees of freedom (it is a topological theory of Schwarz type, \cite{ASS1}). 
Setting the variation of the action equal to zero, we obtain the field
equations:
\beq
dB + [A,B] = D_AB = 0\, ,  \label{eqnB}
\eeq
\beq
dA + \frac{1}{2}[A,  A] = F = 0\,. \,\label{eqnA}
\eeq
The gauge transformations are of the form
\beq
A^g=g^{-1}dg+g^{-1}Ag\,, \hskip 0.3cm B^g=g^{-1}Bg \, .
\eeq	
Since $F$ is the curvature form, Eq.~(\ref{eqnA}) states that the connection $A$ is flat. 
It is this feature that  we shall investigate below  in the context of quantum gauge theory.

\subsection{Feynman diagrams}

\begin{figure}[h]
    \centering
    \setlength{\unitlength}{1cm}   
    \includegraphics[width=7cm]{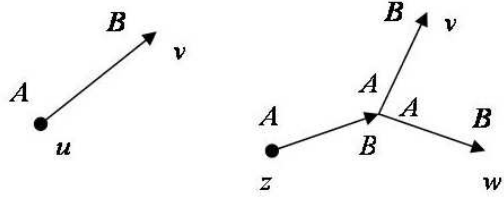}
    \caption{\small\it Diagram building blocks: (a) single edge; (b) vertex.}
    \label{elements.eps}
\end{figure}

It is convenient to rewrite the classical action in the form
\beq
S_{BF} = \tr \int \left( B dA + \frac{1}{2}B [A, A]\right),
\eeq
where the first term can be viewed as a free part of the action (in fact, it corresponds to an Abelian BF theory) while the second term represents the interaction. 
Feynman diagrams in this theory are built of oriented edges pointing from $A$ to $B$ 
and of trivalent vertices with one incoming $B$-field and two outgoing $A$-fields, see Fig. \ref{elements.eps}.

Depending on the choice of $\M$, the propagator corresponding to an oriented edge can be chosen in various ways. 
For the BF theory on a plane, one can choose 
$$
\langle A_a(u) B_b(v) \rangle= \frac{\delta_{ab}}{2\pi} \, d  \arg(u-v) \, ,
$$
where $u$ and $v$ are complex coordinates on the plane, and the right hand side is viewed as a 1-form with respect to $u$.
Note that the choice of propagator corresponds to a particular gauge fixing in the theory. The triple vertex corresponds
to structure constants $f_{abc}$ of the Lie algebra $\GG$.

\begin{figure}[h!]
    \centering
    \setlength{\unitlength}{1cm}   
    \includegraphics[width=10cm]{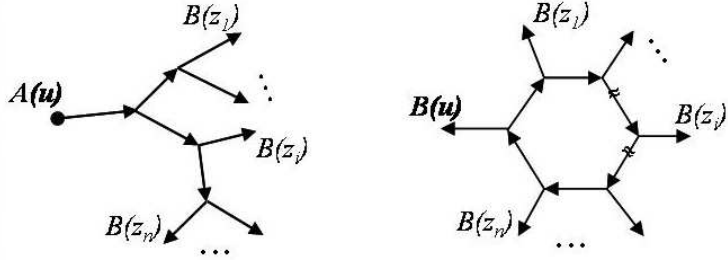}
    \caption{\small\it Basic diagrams: (a) Tree-type diagram, {\rm [T]}\/; (b) Wheel-type diagram, {\rm [W]}\/.}
    \label{treesandwheels.eps}
\end{figure}

Connected Feynman graphs of the BF theory are tree diagrams with one external $A$-field and an arbitrary number of $B$-fields 
(see Fig. \ref{treesandwheels.eps}(a)), and one-loop (or wheel-type)
diagrams with only $B$-fields on the external lines (see Fig. \ref{treesandwheels.eps}(b)). 

\subsection{BF theory with sources}

We shall be interested in the  BF theory with source terms for $B$-field added. 
For the classical action, we have
\beq\label{Seta}
S_\eta = \tr \left( \int_\M BF +  \sum_{i=1}^n \eta_i B(z_i) \right) \,,
\eeq
where we added  classical sources $\eta_i$ at $n$ fixed points, $(z_1, \dots, z_n)$. The partition function is then given by
\beq\label{Keta}
K_\eta (z_1, \dots, z_n) = 
\int e \,\raisebox{10pt}{$S_\eta$} = \int e \, \raisebox{10pt}{$S_{BF} + \dsum_{i=1}^n 
\tr( \eta_i B(z_i))$} \, ,
\eeq
and it can be viewed as a correlation function of the operators  $\exp \tr( \eta_i B(z_i))$ in the theory without sources,
\beq
K_\eta(z_1, \dots,  z_n)=\left\langle \, e^{\tr( \eta_i B(z_i))} \dots
e^{\tr( \eta_i B(z_i))} \right\rangle \, .
\eeq
For an operator $\Oo$, the expectation value  is defined by formula
\beq\label{expectation}
\langle \Oo \rangle_\eta = \left(\dint \Oo e^{S_\eta} \right) \Big/ \left( \dint e^{S_\eta} \right)\,.
\eeq
Thus, 
\beq\label{etaexpect}
\langle \Oo \rangle_\eta= \dfrac{\Big\langle \Oo\, e \, \raisebox{12pt}
{$\sum\limits_{i=1}^n \tr\big(\eta_i B(z_i)\big)$}\Big\rangle}{\Big\langle e\, \raisebox{12pt}{$\sum\limits_{i=1}^n \tr\big(\eta_i B(z_i)\big)$}\Big\rangle} \, \, .
\eeq

In particular, we shall study two cases: when $\Oo$ is the gauge field $A(u)$ and when $\Oo$ is the $B$-field $B(u)$. Note that these are not gauge invariant observables, and that the source terms explicitly break the gauge invariance of the action. 

\begin{figure}[h!]
    \centering
    \setlength{\unitlength}{1cm}   
    \includegraphics[width=5.5cm]{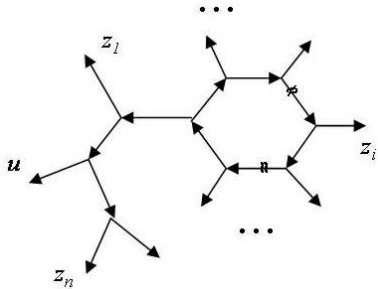}
    \caption{\small\it A typical $B$-field diagram -- a \/{\rm [TW]} composition.}
    \label{treewheel.eps}
\end{figure}

First, we observe that the expectation value of the $A$-field obtains contributions only from tree-type diagrams.
This defines the quantum gauge field $\A$, 
\beq
\A(u)  =  \langle A(u) \rangle_\eta = \,\,\sum_
{\mbox{{\footnotesize\it all trees}}}
\,\Big(\mbox{Fig.~\ref{treesandwheels.eps}(a)}\Big) \,\, .
\label{Aquant} 
\eeq
For a $B$-field, it is slightly more complicated: we obtain all possible wheel-type diagrams hanging on a branch of a tree-type diagram,  see Fig.~\ref{treewheel.eps}. 
\beq
\B(u) =  \langle B(u) \rangle_\eta  =  \sum_
{\mbox{{\footnotesize\it all {\rm [TW]}  compositions}}} 
\Big(\mbox{Fig.~\ref{treewheel.eps}}\Big)\label{Bquant} \\[-4pt]
\eeq

Note that both trees and wheels may have arbitrary lengths, and this is taken into account  in the infinite sums of (\ref{Aquant}) and (\ref{Bquant}). In particular, among tree diagrams  there are short trees (containing only one edge,  $\rm [T(l \!=\!1)\rm]$), see Fig.~\ref{elements.eps}. It is convenient to rewrite Eq.~(\ref{Aquant}) as a sum of two terms
\beq\label{AQsingreg}
\A(u) = \sum_{i=1}^n \eta_i\,d\,\arg(u-z_i) + a(u; z_1, \dots, z_n),
\eeq
where $a(u;  z_1, \dots, z_n)$ is the sum over all trees with length $l>1$, $\rm [T(l \!>\!1)\rm]$.

\subsection{Quantum equations of motion}

We aim at obtaining quantum equations of motion for the BF theory with sources. The canonical way of doing it is by applying the BRST  technique, or rather its generalization --- the Batalin--Vilkovisky method, as the BRST operator does not provide a well defined cohomology needed to define physical observables of the theory. This method implies introducting  ghosts and anti-fields with complimentary ghost numbers and degrees (see, e.g. \cite{BlauAnnals}). We shall instead make use of the graphical representation of the quantum fields --- Eqs. (\ref{Aquant}), (\ref{Bquant}), resp. Fig.~\ref{treesandwheels.eps}, Fig.~\ref{treewheel.eps}, where all terms in the field expansions are present, thus the equations obtained should account for all quantum corrections, including those coming from the gauge-fixing terms.  

\begin{figure}[h!]
    \centering
    \setlength{\unitlength}{1cm}   
    \includegraphics[width=12cm]{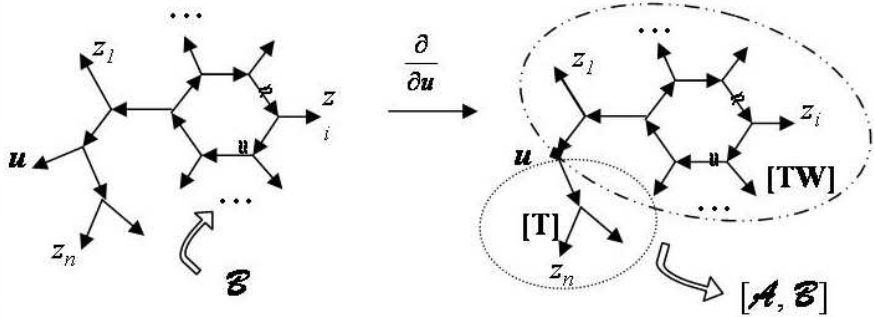}
    \caption{\small\it Equation of motion for $\B$-field.}
    \label{dB.eps}
\end{figure}

On Fig.~\ref{dB.eps} we show the differential of the quantum $\B$-field. By taking the derivative with respect to  the root-point $u$, the corresponding diagram splits into two subgraphs. The first subgraph is a wheel-type diagram, and the second subgraph is a tree. Two subgraphs are related by a Lie bracket corresponding to the vertex where they meet.
Thus, the quantum equation of motion for  $\B$ reads
\beq\label{dBeq}
d \B =- [\A, \B]\, .
\eeq
In fact, it coincides with the classical equation of motion, Eq.~(\ref{eqnB}).

\begin{figure}[h!]
    \centering
    \setlength{\unitlength}{1cm} 
    \includegraphics[width=8.5cm]{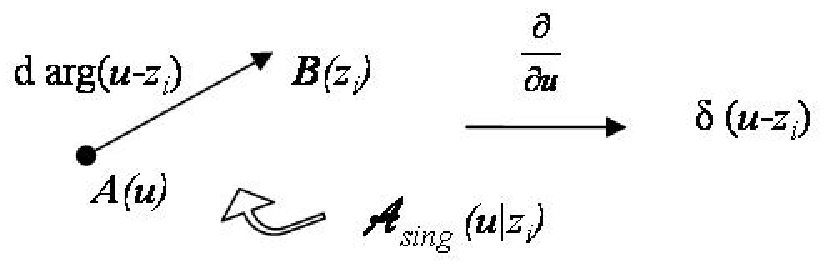}
    \caption{\small\it Equation of motion for $\A$: singular terms.}
    \label{dAsing.eps}
\end{figure}

\begin{figure}[h!]
    \centering
    \setlength{\unitlength}{1cm}   
    \includegraphics[width=12.5cm]{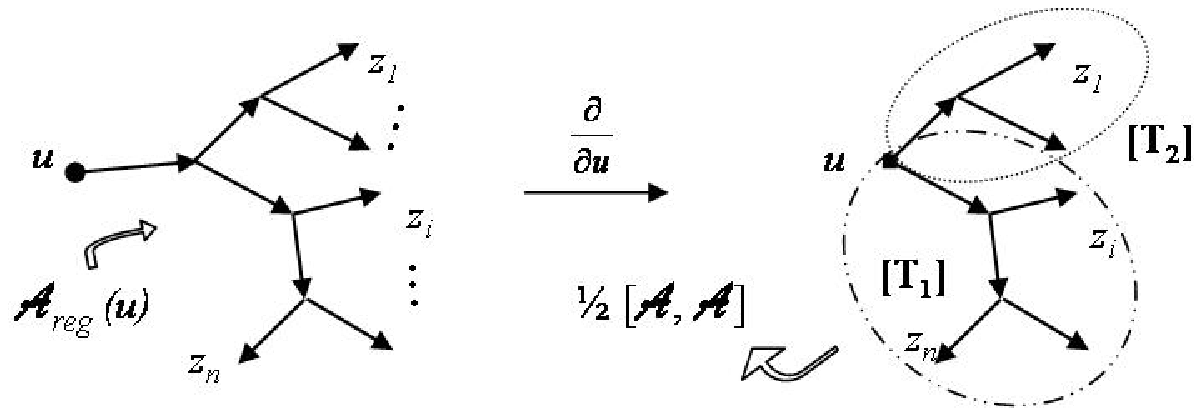}
    \caption{\small\it Equation of motion for $\A$: regular terms.}
    \label{dAreg.eps}
\end{figure}

For the differential of the quantum gauge field $\A$, we use the splitting (\ref{AQsingreg})  to obtain the singular and the regular  parts of the result.  The singular part (one-edge graphs) generates a sum-over-sources term, Fig. \ref{dAsing.eps}. As seen from Fig. \ref{dAreg.eps}, the derivative of the regular part, similarly to the case of the $B$-field, splits into two tree-type subgraphs rooted at $u$.

Thus, the quantum equation for $\A$  takes the form
\beq\label{QAflat}
d \A = - \frac{1}{2} \,[\A, \, \A] +  \sum_{i=1}^n \eta_i\,\delta (u-z_i)\, ,
\eeq
which is again of the same form as the corresponding classical equation of motion.

\bigskip

\section{Equations for correlators and quantum flat connection}

In this Section, we give a physical interpretation of the equations for correlation functions constructed in \cite{To1}. 
These equations fit into a flat connection studied in a more mathematical framework in \cite{AT0906}.

For this purpose, we shall investigate  the dependence of the generating functional of the $B$-field  correlators $K_\eta(z_1,\dots,z_n)$ on the positions of the sources $z_1, \dots, z_n$. That is, we will be interested in the derivatives of the quantum fields $\A$ and $\B$ with respect to coordinates $z_i$.  

Note that the quantum field (\ref{AQsingreg}) is singular at the points where the sources are placed. In order to regularize this singularity,
it is convenient to introduce for each $i$ a new splitting of $\A(u)$ in the form
\beq\label{aiz}
\A_{(i)}(u) =   \frac{\eta_i}{2 \pi}\,d\,\arg(u-z_i) + \A_{(i)}^{reg}(u),
\eeq
where all the unit-length trees but one (connecting the points $u$ and $z_i$) are now kept in the regular part: 
\beqa\label{aireg}
 \A_{(i)}^{reg}(u) & = & \sum_{j\not= i} [T(l\!=\!1); \{u, z_j\}] + \sum_{\mbox{{\footnotesize\it all trees, $l>$1}}}[T]\nonumber\\[8pt]
& = &  \sum_{j\not= i} \frac{\eta_j}{2 \pi}\,d\,\arg(u-z_j) +   a(u; z_1, \dots, z_n)\,.
\eeqa

Observe, that $\A_{(i)}^{reg}(u)$ has no singularity at  $u=z_i$. Let us denote its
value by
\beq\label{aizi}
a_i := \A_{(i)}^{reg}(u; z_1, \dots, z_i, \dots,  z_n)\Big\vert_{u=z_i} .
\eeq

The quantum equation of motion for the $\B$-field leads to the following relation: 
\beqan
d \, {\rm tr} \, \big(  \eta \B(u) \big) & = &
- {\rm tr} \, \big( \eta [\A(u), \B(u)]\big) =
- {\rm tr} \, \big([\eta, \A(u)] \B(u) \big) \\[6pt]
& = &
- {\rm tr} \, [\eta, \A(u)] \, \frac{\partial}{\partial \eta} \, {\rm tr} \, \big( \eta \B(u)\big) \, .
\eeqan
Naively, we should expect the following equation for $K_\eta (z_1, \dots, z_n)$ to hold:
\beq\label{formal_eq_K}
d_{z_i} K_\eta (z_1, \dots, z_n) + {\rm tr} \, 
[\eta_i, \A(z_i)] \, \frac{\partial}{\partial \eta_i}  \,   K_\eta (z_1, \dots, z_n) =0 \, .
\eeq
Here $d_{z_i}$ stands for the de Rham differential with respect to the coordinate $z_i$
(note that it includes both holomorphic and anti-holomorphic differentials). 
Since $\A(z_i)$ is ill-defined, we need to re-examine the Feynman graphs which contribute in the right hand side of Eq.(\ref{formal_eq_K}).

\begin{figure}[!h]
    \centering
    \setlength{\unitlength}{1cm}  
    \includegraphics[width=6.5cm]{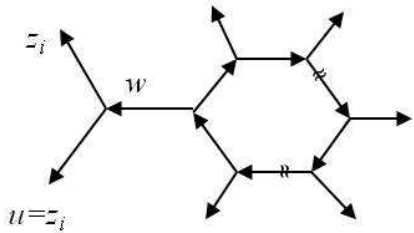}
    \caption{\small\it Vanishing $B$-field diagram.}
    \label{Kgraph.eps}
\end{figure}

The only interesting (different from the naive approach) case is the diagram shown on Fig. \ref{Kgraph.eps}. Its contribution vanishes because of the  factor $(d \arg (w-z_i))^2=0$ in the integrand of the corresponding Feynman integral. Hence, the one-edge tree connecting $w$ and $z_i$ does not contribute in the derivative of $K_\eta$, and the renormalized quantum formula
replacing Eq.(\ref{formal_eq_K}) is 

\beq\label{dK}
d_{z_i} K_\eta +{\rm tr} \, 
[\eta_i, a_i] \, \frac{\partial}{\partial \eta_i}  \,   K_\eta =0 \, .
\eeq

Equations (\ref{dK}) for different $i$ can be put together in one equation
\beq \label{dK2}
 d K_\eta + {\rm tr} \, \sum_{i=1}^n
[\eta_i, a_i] \, \frac{\partial}{\partial \eta_i}  \,   K_\eta =0 \, ,
\eeq
where $d$ is the total de Rham differential for all variables $z_1, \dots, z_n$.
For functions $\alpha_i(\eta_1, \dots, \eta_n) \in \GG, i=1,\dots, n$, operators 
\beq
D_\alpha= {\rm tr}\, \sum_{i=1}^n \, [\eta_i, \alpha_i] \,
\frac{\partial}{\partial \eta_i}
\eeq
form an interesting Lie algebra
\beq
[D_\alpha, D_\beta] = D_{\{ \alpha, \beta\} } \, 
\eeq
with Lie bracket
\beq{\LARGE }
\{ \alpha, \beta\}_i =D_\alpha \beta_i - D_\beta \alpha_i + [\alpha_i, \beta_i]. 
\eeq
One can view the collection of 1-forms $(a_1, \dots, a_n)$ as components
of a connection ${\mathbb A} =(a_1, \dots, a_n)$ with values in this Lie algebra.
Then, equation (\ref{dK2}) for correlation functions simply reads
$$
d K_\eta + D_{{\mathbb{A}}} K_\eta =0\, .
$$

\begin{figure}[ht]
    \centering
    \includegraphics[width=13.5cm]{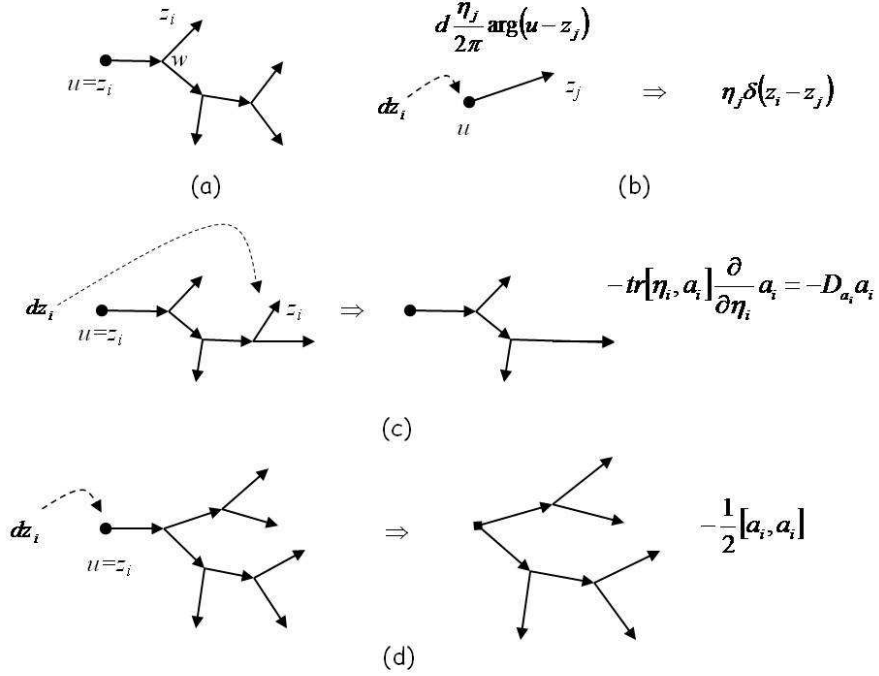}
    \caption{\small\it Graphic $z_i$ differentiation of $a_i$.}
    \label{ai_graphs}
\end{figure}

Similarly, for the differential 
of gauge field $\A(u)$ with respect to the source positions we obtain
\beq \label{d_iA}
d_{z_i} \A(u) =- {\rm tr} \, [\eta_i, a_i] \, \frac{\partial}{\partial \eta_i} \, \A(u) =
- D_{a_i} \A(u).
\eeq
Note that for $j \neq i$ we can replace $\A(u)$ by $\A_{(j)}^{reg}$. Indeed, the 
one-edge tree which is subtracted from $\A(u)$ to get  $\A_{(j)}^{reg}$  (the edge connecting $u$ to $z_j$) does not
contribute neither to the left hand side nor to the right hand side of
Eq.(\ref{d_iA}). Then,  putting $u=z_j$ yields
\beq
d_{z_i} a_j = - D_{a_i} a_j.
\eeq

We will now show that the curvature ${\mathbb F}$ of ${\mathbb A}$ vanishes \cite{AT0906}.
The curvature is defined as
\beq
{\mathbb F}= d{\mathbb A} + \frac{1}{2} \, \{ {\mathbb A}, {\mathbb A} \} \, .
\eeq
We will first compute its components ${\mathbb F}_{ij}$ corresponding to 
two different coordinates $z_i \neq z_j$ (note that the curvature has
holomorphic, anti-holomorphic and mixed components). 
The curvature ${\mathbb F}_{ij}$ has $n$ components $({\mathbb F}_{ij})_k$
for $k=1, \dots, n$. The components with $k \neq i,j$ vanish identically.
For the remaining components, we have
\beq
({\mathbb F}_{ij})_i=d_{z_j} a_i + D_{a_j} a_i=0 \, , \hskip 0.5cm
({\mathbb F}_{ij})_j=d_{z_i} a_j + D_{a_i} a_j=0.
\eeq

The curvature $F_{ii}$ has only one nonvanishing component,
\beq\label{Fii}
({\mathbb F}_{ii})_i = d_{z_i} a_i + D_{a_i} a_i + \frac{1}{2} [a_i, a_i]\, .
\eeq
In more detail, put $a_i = \alpha_i dz_i + \bar{\alpha}_i d\bar{z}_i$ to obtain
\beq
({\mathbb F}_{ii})_i = \partial_{z_i} \bar{\alpha}_i - \bar{\partial}_{z_i} \alpha_i 
+ D_{\alpha_i} \bar{\alpha_i} - D_{\bar{\alpha}_i} \alpha_i + [\alpha_i, \bar{\alpha}_i].
\eeq
In order to compute this expression, we consider the differential $d_{z_i} a_i $.There are several types of diagrams which contribute  (see Fig. \ref{ai_graphs}). Note that graphs of  type (a) vanish, as in the derivative of $K_\eta$. Graphs of types  (b) and (c) generate source terms and covariant derivative terms. Graphs of type (d) accounts for an extra $z_i$ dependence due to the root of the tree. The result is
\beq\label{daizi}
d_{z_i}a_i(z_i) + D_{a_i} a_i + \frac{1}{2}\,[a_i, a_i] = \sum_{j\not=i} \eta_j\, \delta(z_i-z_j) \,.
\eeq

That is, away from the sources positions, the connection is flat,
\beq
d{\mathbb A} + \frac{1}{2} \, \{ {\mathbb A}, {\mathbb A} \} =0.
\eeq
With sources taken into account, we have 
${\mathbb F}=({\mathbb F}_1, \dots , {\mathbb F}_n)$, where
\beq
{\mathbb F}_i = \sum_{j\neq i} \eta_j \delta(z_i - z_j).
\eeq

\bigskip

\section{Outlook}

The Torossian connection discussed in Section 3 is a close relative of the Knizhnik-Zamolodchikov (KZ)
connection in the WZW theory. Recall that the KZ connection describes correlators of primary fields, and 
that it has the form
\beq
d \Psi + \mathbb{A}_{KZ} \, \Psi =0\, , \hskip 0.5cm
\mathbb{A}_{KZ} = \frac{1}{2\pi i} \sum_{i,j} \, t_{i,j} \, d\ln(z_i - z_j),
\eeq
where $t_{i,j} = \sum_a e_a^i \otimes e_a^j$ are operators acting on the product of irreducible 
representation of $\GG$ carried by primary fields placed at the points $z_1, \dots, z_n$. Note that
operators $t_{i,j}$ play the role of one-edge trees, and the propagator has the form 
$d\ln(z_i - z_j)/2\pi i$.

The KZ connection admits the second interesting interpretation: one can view it as an equation
on the wave function of the Chern-Simons topological field theory with $n$ time-like Wilson lines
(corresponding to primary fields) \cite{FrKing}. From this perspective, holonomy matrices
of the flat connection $\mathbb{A}_{KZ}$ correspond to braiding of Wilson lines in the
Chern-Simons theory.

It would be very interesting to find a three dimensional topological field theory which 
has the Torossian connection as an equation on the wave function. Of course, such
a theory must have non-local observables (similar to Wilson lines) which will 
correspond to insertions of operators $\exp({\rm tr}\, \eta_i B(z_i))$ in the
2-dimensional theory.

\bigskip
{\small\it E-mail address:} {\small\texttt anton.alekseev@unige.ch}\\
\indent{\small\it E-mail address:} {\small\texttt nilieval@mail.cern.ch}

\end{document}